\documentclass[preprint,aps]{revtex4}
\DeclareUnicodeCharacter{00A0}{ }
\usepackage{graphicx}
\usepackage{textgreek} 
\usepackage{verbatim}   
\usepackage{amssymb}
\newcommand{\be}{\begin{eqnarray}}
\newcommand{\ee}{\end{eqnarray}}
\usepackage{url}
\usepackage{cancel} 
\newcommand{\nn}{\nonumber}

\usepackage{amsmath,amssymb}
\usepackage{mathtools} 

\begin{document}

\title{Interacting tachyonic scalar field}

  \author{A Kundu${}^1$\footnote{kunduaku339@gmail.com}, S D Pathak${}^1$\footnote{prince.pathak19@gmail.com} and V K Ojha${}^2$\footnote{ohjavikash@gmail.com}}
  
  \affiliation{${}^1$ Department of Physics, Lovely Professional University, Phagwara, Punjab, 144 411, India.\\ ${}^2$School of Sciences, SAGE University Bhopal, Madhya Pradesh, 462 022, India}

 \begin{abstract}
  We discuss the coupling between dark energy and matter by considering a homogeneous tachyonic scalar field as a candidate for dark energy. We obtained the functional form of scale factor by assuming that the coupling strength depends linearly on the Hubble parameter and energy density. We also estimated the cosmic age of the universe for different values of coupling constant.

 \end{abstract}
 
\maketitle
\section{Introduction}
\noindent
In the year 1998, type Ia Supernova observation\cite{Riess_1998,Perlmutter_1999} revealed the accelerated expansion of the universe. To explain the observed accelerated expansion of the universe, the idea of dark energy has been introduced. One of the remarkable features of dark energy is its negative pressure which results in the negative equation of state ($\omega_{\textrm{de}}$). The observed cosmic acceleration is possible only for $\omega_{\textrm{de}}<-1/3$. The cosmological constant is one of the potential candidates of the dark energy bearing equation of state $\omega_{\lambda }=-1$.  The cosmological constant in the constant dark energy model suffers two serious issues namely the coincidence problem and cosmological constant problem. To resolve these two open problems interacting dark energy model has been proposed by a number of authors\cite{Chimento_2010,Chimento_2008,Bertolami_2012,Wang_2007,lu2012investigate,Farajollahi_2012,Zimdahl_2012,Yang_2018,Cao_2011,Di_Valentino_2020,Verma_2012,Verma_2013,V_liviita_2010,Pan_2018,Amendola_2018,B_gu__2019,Pan_2019,Papagiannopoulos_2020,Savastano_2019,von_Marttens_2019,yang2019reconstructing,Asghari_2019} by considering the dynamical behavior of dark energy.

In the interacting dark energy models, transfer of energy between the dominant components of the universe ((matter, dark energy, radiation)) is considered. The interaction term  $Q$ in the interacting dark energy model concern the transfer of energy between the dominant components of the universe. One of the major challenges in the interacting dark energy model is to fix the exact functional form of coupling term $Q$. Different functional forms of coupling term have been used over the years. \cite{Chimento_2010,Chimento_2008,Bertolami_2012,Wang_2007,lu2012investigate,Farajollahi_2012,Zimdahl_2012,Yang_2018,Di_Valentino_2020}

In the absence of a fundamental theory of dark sector physics, the choice of coupling strength in interacting dark energy model might be purely phenomenological.  One of the motivations to fix the form of coupling term comes from the dimensional argumentativeness applied to the continuity equation of energy for the components of the universe. The form of coupling strength can be either the linear function of energy density and the Hubble parameter of the field or the function of the first time derivative of energy density. 

In this article, we have discussed the scaling solution of energy density for the two dominant interacting components of the universe. One component is the scalar field $\phi$ which we take as a candidate of dynamical dark energy and the other is dust matter.
These two components are interacting via the transfer of their energy into each other. Further, we have estimated the age of the universe and its variation with the varying coupling constant.


\section{Background theory and motivation}
\label{intro}
\noindent
In this section, we present a brief introduction to the working background theory of interacting dark energy model. The FLWR metric for flat($K=0$) universe given as
\begin{equation}
\textrm{d}s^{2}=-\textrm{d}t^{2}+a^2\left(t\right) \left( \textrm{d}x^{i}\right) ^{2}\label{eq:metric},
\end{equation}
where $a(t)$ is the expansion scale factor and $i=1,2,3$ represents the spatial component of spacetime. The Friedman equation can be obtained by solving the Einstien field equation for the above metric 
%
\begin{equation}
H^{2}=\frac{8\pi G}{3}(\rho _{m} + \rho_{\textrm{de}})\label{eq:friedmann-eqn},
\end{equation}
where $\rho_{\textrm{m}}$, $\rho_{\textrm{de}}$ are energy density of dust matter and dark energy respectively. 
We are considering a dynamical dark energy model. Although cosmological constant is a potential candidate of constant dark energy, we assume the scalar field as a candidate for dynamical dark energy. A number of scalar fields (quintessence, phantom, tachyonic, etc.) have been introduced in physics in different contexts. We scrutinize the dynamical behavior of the tachyonic scalar field in the interacting dark energy model by assuming it as one of the candidates for dynamical dark energy.

The Lagrangian of tachyonic scalar field appears in string theory in the formulation of tachyon condensate\cite{Sen_2002,Sen1_2002,Sen2_2002} given as
\begin{equation}
\mathcal{L}=-V\left( \phi \right) \sqrt{1-\partial ^{\mu }\phi\partial _{\mu }\phi }\label{eq:lagrangian-tachyon},\nn
\end{equation}
where $V(\phi)$ is the potential of the field. The equation of motion  for the spatially homogeneous tachyonic scalar field can be written as 
\begin{equation}\label{EoM}
\frac{\ddot{\phi}}{1-\dot{\phi^2}}+3H\dot{\phi }+\frac{V'(\phi)}{V(\phi)}=0, \nn
\end{equation}
and the energy-momentum tensor is
\begin{equation}\label{emt}
T^{\mu \nu }=\frac{\partial \mathcal{L}}{\partial \left( \partial _{\mu }\phi \right) }\partial ^{\nu }\phi -g^{\mu \nu }\mathcal{L}.
\end{equation}
From the energy-momentum tensor Eq(\ref{emt}), the $T^{00}$ component gives the energy density while the $T^{11}$ component leads to pressure for the tachyonic scalar field, i.e.
\begin{equation}
\rho =\frac{V\left( \phi \right) }{\sqrt{1-\partial ^{\mu }\phi \partial _{\mu }\phi}},\label{eq:energy-density} \nn
\end{equation}
and
\begin{equation}
p=-V\left( \phi \right) \sqrt{1-\partial ^{\mu }\phi \partial _{\mu }\phi}\label{eq:pressure}. \nn
\end{equation}
For the spatially homogeneous tachyonic scalar field, the energy density and the pressure reduces to the following form
\begin{equation}
\rho =\frac{V\left( \phi \right) }{\sqrt{1-\dot{\phi^2}}},\;\;\; p=-V\left( \phi \right) \sqrt{1-\dot{\phi^2}}.\label{eq:spatially-homogeneous-p-rho} \nn
\end{equation}


\section{Interaction between the components}
We consider the two dominant components of the universe are spatially homogeneous tachyonic scalar field (as a candidate of dynamical dark energy) and the dust matter. These two components interact via the transfer of energy. The continuity equation leads to conservation of energy for the two-components. For non-interacting case as, we have
\begin{equation}\label{conser1}
\dot{\rho }_{\phi }+3H\left( 1+\omega _{\phi}\right) \rho_{\phi } = 0,\nn
\end{equation}
and 
\begin{equation}\label{conser2}
\dot{\rho }_{m}+3H\left( 1+\omega _{m}\right) \rho _{m} =0, \nn
\end{equation}
where $w_\phi = p_\phi /\rho_\phi
$. Here $
w_\phi$, $w_m$ is the equation of state parameter for the scalar field $\phi
$ and the dust matter respectively.
During the interaction, individual components can violate the energy conservation but the total energy is conserved. 
Thus, the energy conservation with the interaction term $Q$ can be represented by the following equations
\begin{align}
&\dot{\rho }_{\phi }+3H\left( 1+\omega _{\phi}\right) \rho_{\phi } =-Q\nonumber,\\\label{eq:conser3}\\
&\dot{\rho }_{m}+3H\left( 1+\omega _{m}\right) \rho _{m} = Q.\nonumber
\end{align}
In the absence of the fundamental theory of the dark sector, there are some possible linear and non linear functional forms of interaction term Q. One can fix the functional form of $Q$ purely based upon the phenomenological argument which is it should be  dimensionally compatible with the left-hand side of Eq(\ref{eq:conser3}). Thus, it is natural that the coupling parameter $Q$ should be a function of energy density and Hubble parameter or the rate of change of energy density of components. Based on phenomenological motivation, several authors\cite{pathak2016thermodynamics,Amendola_2003,Berger_2008,Verma_2014,Shahalam_2015} proposed different forms of interaction term in the interacting dark energy model. 

In our model we considered a specific functional form of coupling term which is linearly proportional to Hubble parameter $H$ as well as the energy density $\rho_{\phi}$ of the tachyonic scalar field. This form of coupling term has been used in the literature\cite{Pav_n_2008}. Thus we have the following form for interaction term

\begin{equation}
Q=\alpha H\rho_\phi\label{eq:coupling-expression},
\end{equation}
where $\alpha$ is the proportionality constant.
One can solve Eq(\ref{eq:conser3}) for given interaction strength Eq(\ref{eq:coupling-expression}) to obtain the following scaling solution of the energy density of components\cite{Akash_2020}

{\begin{equation}\label{sol1}
\frac{\rho _{\phi }}{\rho _{\phi }^{0}} =\left( \frac{a}{a_{0}}\right)^{-\beta},\nn
\end{equation}}
and 
{\begin{equation}\label{sol2}
\frac{\rho _{m}}{\rho_{m}^{0}}=\left( \frac{a}{a_{0}}\right)^{-3\left( 1+\omega_{m}\right) }+\frac{\beta \rho _{\phi }^{0}}{\rho_{m}^{0}\left[ \beta -3\left( 1+\omega_{m}\right) \right] }\left[\left( \frac{a}{a_{0}}\right)^{-3\left( 1+\omega_{m}\right) }-\left( \frac{a}{a_{0}}\right) ^{-\beta }\right], \nn
\end{equation}
where $\beta=\alpha+3(1+\omega_\phi)$},  $\rho_\phi^0$  ($\rho_m^0$) is the present value of scalar field (dust matter), and $a_0$ is the present value of scale factor.
If $\dot{\phi}\rightarrow 0$, i.e. for constant $\phi$ we get  $\rho_{\phi }\rightarrow\rho_{\lambda }~\textrm{(some constant)}$. In this approximation limit tachyonic scalar field (TSF) mimics the cosmological constant, and we have $\beta=\alpha$. Hence, the scaling solution reduces to the following form
\begin{equation}\label{eq:scaling-1-with-c-phi-tendsto0}
\frac{\rho _{\lambda }}{\rho^{0}_\lambda}=\left(\frac{a}{a_{0}}\right)^{-\alpha},
\end{equation}

\begin{equation}
\frac{\rho_{m}}{\rho_{m}^{0}} = \left( \frac{a}{a_{0}}\right)^{-3\left( 1+\omega_{m}\right) }+\frac{\alpha \rho_{\lambda }^{0}}{\rho^{0}_{m} \left[ \alpha -3\left( 1+\omega_{m}\right) \right] }\left[\left( \frac{a}{a_{0}}\right)^{-3\left( 1+\omega_{m}\right) }-\left( \frac{a}{a_{0}}\right)^{-\alpha }\right].\label{eq:scaling-2-with-c-phi-tendsto0}
\end{equation}


\section{Functional form of scale factor}
\noindent
In this section, we investigate the evolution of the expansion scale factor ($a$) for the two important cases. The interaction proportionality constant term ($\alpha$) can be either zero or non zero. So, we are considering the following two case

\subsection{For $\alpha=0$}
From the Friedmann equations (Eq(\ref{eq:friedmann-eqn})) along with the scaling solutions Eq(\ref{eq:scaling-1-with-c-phi-tendsto0}) and Eq(\ref{eq:scaling-2-with-c-phi-tendsto0}),  we have
\begin{equation}
H=\sqrt{\frac{8\pi G}{3}}\sqrt{\rho_\lambda^0+\rho^{0}_{m}x^{-3(1+\omega_m)}}\label{eq:hubble-expression-without-c},
\end{equation}
where $x=\dfrac{a}{a_{0}}$, $a_{0}$ is the present value scale factor.
Equation (\ref{eq:hubble-expression-without-c}) gives 
\begin{equation}
\textrm{d}t=\frac{A\textrm{d}x}{x\sqrt{1+B^2x^{-3(1+\omega_m)}}}\label{eq:differential-without-c},
\end{equation}
where
\begin{equation}\label{A}
A=\sqrt{\frac{3}{k\rho_\lambda^0}} ~~,~~ B=\sqrt{\frac{\rho_m^0}{\rho_\lambda^0}} ~~,~\mathrm{and}~~ k=8\pi G\nn
\end{equation} 
Solving Eq(\ref{eq:differential-without-c}), we get the analytical expression for $x$ as a function of cosmic time $t$
\begin{equation}
x=\left[B^{-2}\left(\text{tanh}\left(\frac{3(1+\omega_m)t}{2A}\right)\right)^2-B^{-2}\right]^{-\frac{1}{3(1+\omega_m)}}.\label{eq:x-without-c}
\end{equation}
\subsection{For $\alpha\ne0:$}
In this case, Eq((\ref{eq:hubble-expression-without-c})) leads to the following equation  
\begin{equation}\label{eq:differential-with-c}
\textrm{d}t=\frac{A'\textrm{d}x}{x^{(1-\frac{\alpha}{2})}\sqrt{1+B'^2x^{\alpha-3(1+\omega_m)}}},
\end{equation}
where
\begin{equation}
A'=\sqrt{\frac{3(1+\omega_m)-\alpha}{k\rho_\lambda^0(1+\omega_m)}}~~,~~\textrm{and}~~ B'=\sqrt{\frac{3(1+\omega_m)\rho_m^0-\alpha(\rho_\lambda^0+\rho_m^0)}{3(1+\omega_m)\rho_\lambda^0}}. \nn
\end{equation}
Solving Eq(\ref{eq:differential-with-c}) to obtain an expression of $t$ in terms of $x$, we have
\begin{equation}
t=\frac{2A'}{(-1)^{n} B'^{\alpha'}[\alpha-3(1+\omega_m)]}\left[\sqrt{1+B'^2x^{\alpha-3(1+\omega_m)}}\;_2\textrm{F}_1\left(\frac{1}{2},n,\frac{3}{2},\left(\sqrt{1+B'^2x^{\alpha-3(1+\omega_m)}}\right)^2\right)\right]\label{eq:hypergeometric},
\end{equation}
where
\begin{equation}
\alpha'=\dfrac{\alpha}{\alpha-3(1+\omega_{m})},\;\;\;\;\;\;\;\; n=\frac{\frac{\alpha}{2}-3(1+\omega_m)}{\alpha-3(1+\omega_m)}\label{eq:n}.
\end{equation}
The function $_2\textrm{F}_1$ hypergeometric functional series defined by \cite{Hypergeometric_book}
	\begin{equation}
	_2\textrm{F}_1(a,b,c,z)=\sum_{m=0}^{\infty}\frac{(a)_m(b)_m}{(c)_m}\frac{z^m}{m!}\label{eq:hypergeo-series}, 
	\end{equation}
	where
	\begin{equation}
	(a)_m= \left\{
	\begin{array}{ll}
	1 &~ m=0,\\
	a(a+1)..(a+m-1) &~ m>0.
	\end{array} \right. \nn
	\end{equation}
By using the definition of $_2\textrm{F}_1$, Eq(\ref{eq:hypergeometric}) can be rewritten as\\
\begin{multline}
t=\dfrac{2A'}{(-1)^nB'^{\alpha'}\left[\alpha-3(1+\omega_m)\right]}\left(\sqrt{1+B'^2x^{\alpha-3(1+\omega_m)}}+\dfrac{1}{3}n\dfrac{(\sqrt{1+B'^2x^{\alpha-3(1+\omega_m)}})^3}{1!}\right.\\ 
+\left.\dfrac{1}{5}n(n+1)\dfrac{(\sqrt{1+B'^2x^{\alpha-3(1+\omega_m)}})^5}{2!}+...\right)\label{eq:t-with-c}.
\end{multline}
When $\alpha=0$ (from Eq.(\ref{eq:n})), we have $n=1$, $\alpha'=0$, $A'=A$ and, $B'=B$. Hence, the term inside the bracket of Eq.(\ref{eq:t-with-c}) can be written as
\begin{multline}
\text{tanh}^{-1}\sqrt{1+B^2x^{-3(1+\omega_m)}}=\sqrt{1+B^2x^{-3(1+\omega_m)}}+\dfrac{1}{3}(\sqrt{1+B^2x^{-3(1+\omega_m)}})^3,\\
+\dfrac{1}{5}(\sqrt{1+B^2x^{-3(1+\omega_m)}})^5+...\label{eq:series-arctanh}.
\end{multline}
Using Eq(\ref{eq:series-arctanh}) and Eq(\ref{eq:t-with-c}) we get
\begin{equation}
t=\frac{2A}{3(1+\omega_m)}\text{tanh}^{-1}\sqrt{1+B^2x^{-3(1+\omega_m)}}\label{eq:t'}.
\end{equation}
In the limit $\alpha\longrightarrow0$, Eq(\ref{eq:t-with-c}) reduces to the same form which we will obtain by integrating the Eq (\ref{eq:differential-without-c}).



\section{Age of the universe (AoU)}
The age of the universe has been discussed in the Ref \cite{Verma_2013} by considering two-phase (interacting and non-interacting dark energy) evolution of the universe. Recent data\cite{planck2020} provide the present value of Hubble parameter $H_0$ to be approximately $67.66\pm0.42$ km/s/Mpc and the present value of normalized energy density parameters are $\Omega_{\lambda}^0=0.6889\pm0.0056$, $\Omega_{m}^0=0.3111\pm0.0056$. In this section, we estimate the age of the universe with and without coupling between the matter and the tachyonic scalar field (candidate of dynamical dark energy). 


\subsection{Without interaction ($\alpha=0$)}
One can find the cosmic age in the absence of interaction by integrating the Eq (\ref{eq:differential-without-c}) to get
\begin{equation}
t_a=\int_{0}^{1}\frac{A\textrm{d}x}{x\sqrt{1+B^2x^{-3(1+\omega_m)}}}.\label{eq:integration-t-with-x-for-nocoupling} \nn
\end{equation}
For the Hubble constant $H_0=67.66$ kms$^{-1}$Mpc$^{-1}$, energy density parameter $\Omega^0_{\lambda}=0.6889$, and $\Omega^0_{m}=0.3111$, we found the age of the universe to be $t_a\approx 0.9543H_0^{-1}\approx13.52$ Gyr. The value of $H_0^{-1}$ is $\approx14.167~\text{Gyr}$.


\subsection{With interaction ($\alpha\ne0$)}
	In presence of interaction the age of the universe can be derived from Eq(\ref{eq:differential-with-c}) and it comes                   out be \cite{Condon_2018}
	\begin{equation}
	t_a(x)=\int_{0}^{x}\frac{\textrm{d}x'}{x'H(x')}, \nn
	\end{equation}
	which can be modified as
	\begin{equation}
	t_a(x)H_0=\int_{0}^{x}\frac{\textrm{d}x'}{x'E(x')}\label{eq:present-age},
	\end{equation}
	where $E(x)=\dfrac{H(x)}{H_0}$ and, $x=\dfrac{a}{a_0}$. Using Eq(\ref{eq:friedmann-eqn}), Eq(\ref{eq:scaling-1-with-c-phi-tendsto0}), and Eq(\ref{eq:scaling-2-with-c-phi-tendsto0}) for $\omega_m=0$, we have
	\begin{equation}
	E(x)=\sqrt{\left(\Omega^{0}_{m}+\frac{\alpha}{\alpha-3}\Omega^{0}_{\lambda}\right)x^{-3}-\Omega^{0}_{\lambda}\left(\frac{3}{\alpha-3}\right)x^{-\alpha}}\label{eq:Eq-value}.\nn
	\end{equation}
	Thus the Eq(\ref{eq:present-age}) gives
	\begin{equation}
	t_a(x)H_0=\int_{0}^{1}\frac{\textrm{d}x'}{\sqrt{\left(\Omega_{m}^0+\frac{\alpha}{\alpha-3}\Omega_{\lambda}^0\right)(x')^{-1}-\Omega_{\lambda}^0\left(\frac{3}{\alpha-3}\right)(x')^{2-\alpha}}}.\label{eq:lookback-integral}
	\end{equation}
The cosmic age of universe in the presence of coupling $\alpha$ for its various numerical values is (from $\alpha=0$ to $\alpha=0.9$) given in the \textit{Table} [\ref{tab:tab1}] for $\Omega_{\lambda}^0=0.7$, $\Omega_{m}^0=0.3$ and,  $\Omega_{\lambda}^0=0.6889$, $\Omega_{m}^0=0.3111$.

\begin{table}
	\begin{tabular}{ll}
	\qquad\qquad\qquad\qquad\ \ \ \ \\
	\begin{tabular}{|l|l|}
		\hline
		$\alpha$ & \ \ \ \  AOU in terms of $H_0^{-1}$\\
		\hline
		$0$ & \ \ \ \ $0.964$ \\
		$0.1$ & \ \ \ \ $0.981$ \\
		$0.2$ & \ \ \ \ $1.001$ \\
		$0.3$ & \ \ \ \ $1.024$ \\
		$0.4$ & \ \ \ \ $1.052$ \\
		$0.5$ & \ \ \ \ $1.088$ \\
		$0.6$ & \ \ \ \ $1.134$ \\
		$0.7$ & \ \ \ \ $1.199$ \\
		$0.8$ & \ \ \ \ $1.311$ \\
		$0.9$ & \ \ \ \ $2.222$ \\
		\hline
	\end{tabular}
\qquad\qquad\qquad
	\begin{tabular}{|l|l|}
		\hline
		$\alpha$ & \ \ \ \ AOU in terms of $H_0^{-1}$\\
		\hline
		$0$ & \ \ \ \ $0.964$ \\
		$0.1$ & \ \ \ \ $0.970$ \\
		$0.2$ & \ \ \ \ $0.989$ \\
		$0.3$ & \ \ \ \ $1.010$ \\
		$0.4$ & \ \ \ \ $1.036$ \\
		$0.5$ & \ \ \ \ $1.068$ \\
		$0.6$ & \ \ \ \ $1.109$ \\
		$0.7$ & \ \ \ \ $1.166$ \\
		$0.8$ & \ \ \ \ $1.254$ \\
		$0.9$ & \ \ \ \ $1.459$ \\
		\hline
	\end{tabular}
	\end{tabular}
	\caption{Age of Universe (AOU)  $t_a$ for different values of coupling constant $\alpha$. First table (left one) is for $\Omega_{\lambda}^0=0.7, ~ \Omega_{m}^0=0.3$ and second one (right one) is for $\Omega_{\lambda}^0=0.6889$, $\Omega_{m}^0=0.3111$}
\label{tab:tab1}
\end{table}



\section{Variation of $\Omega_{\lambda}$, $\Omega_{m}$ with $\textrm{ln}~a$}
Here we have observed the variation of energy density parameter $\Omega_{\lambda}$ and $\Omega_{m}$ as a function of $\eta=\text{ln}~a$ by following the method of Ref.\cite{Bahamonde_2018}. We can define two new variable $\mathcal{X}$ and $\mathcal{Y}$ such that $\mathcal{X}=\dfrac{8\pi G}{3H^2}\rho_{\lambda}=\Omega_{\lambda}$, and $\mathcal{Y}=\dfrac{8\pi G}{3H^2}\rho_m=\Omega_m$. Differntiating $\mathcal{X}$ and $\mathcal{Y}$ with $\eta$, we have

\begin{figure}[!htp]
\begin{minipage}[c]{1\textwidth}
\small{(a)}\includegraphics[width=7.8cm,height=5cm,clip]{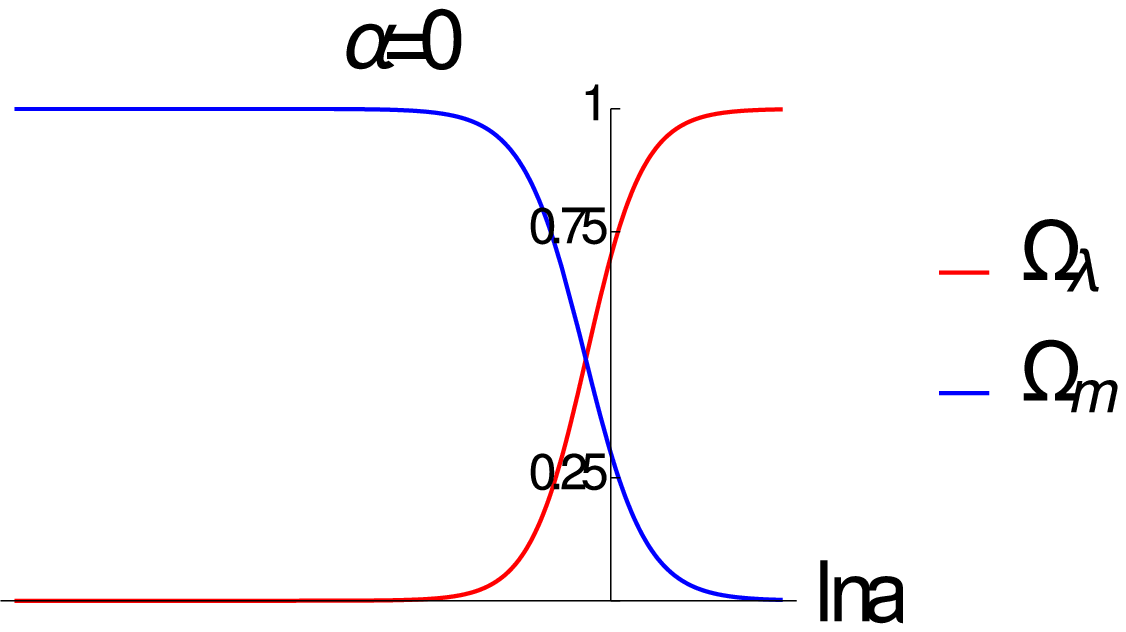}
\hspace{0.1cm}
\small{(b)}\includegraphics[width=7.8cm,height=5cm,clip]{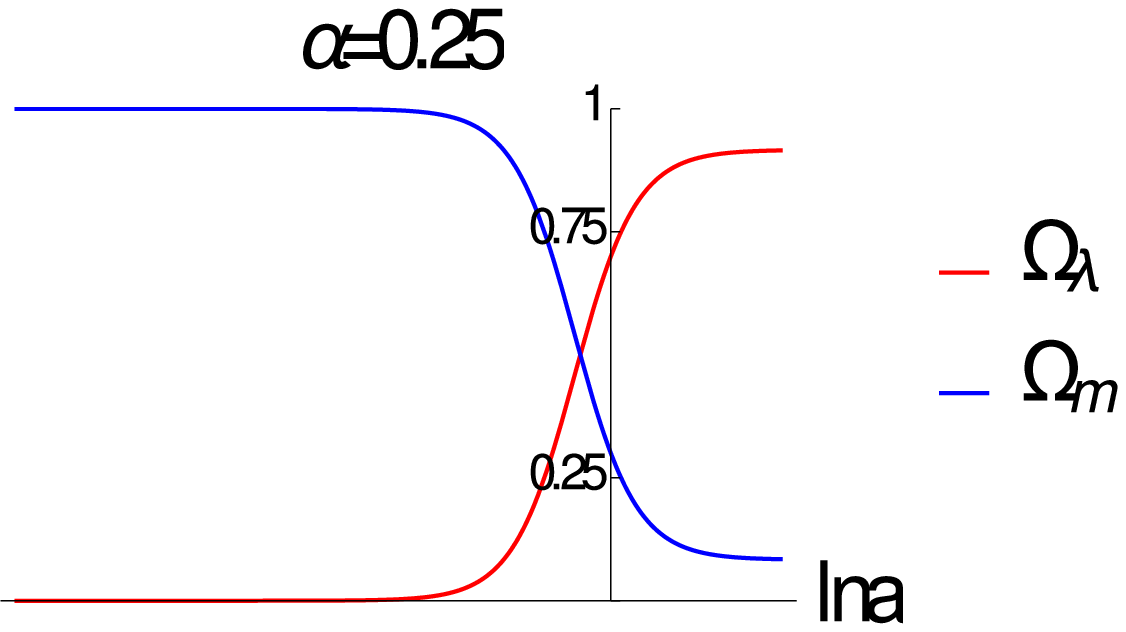}
\vspace{5 mm}
\end{minipage}
\vspace{7 mm}
\begin{minipage}[c]{1\textwidth}
\small{(c)}\includegraphics[width=7.8cm,height=5cm,clip]{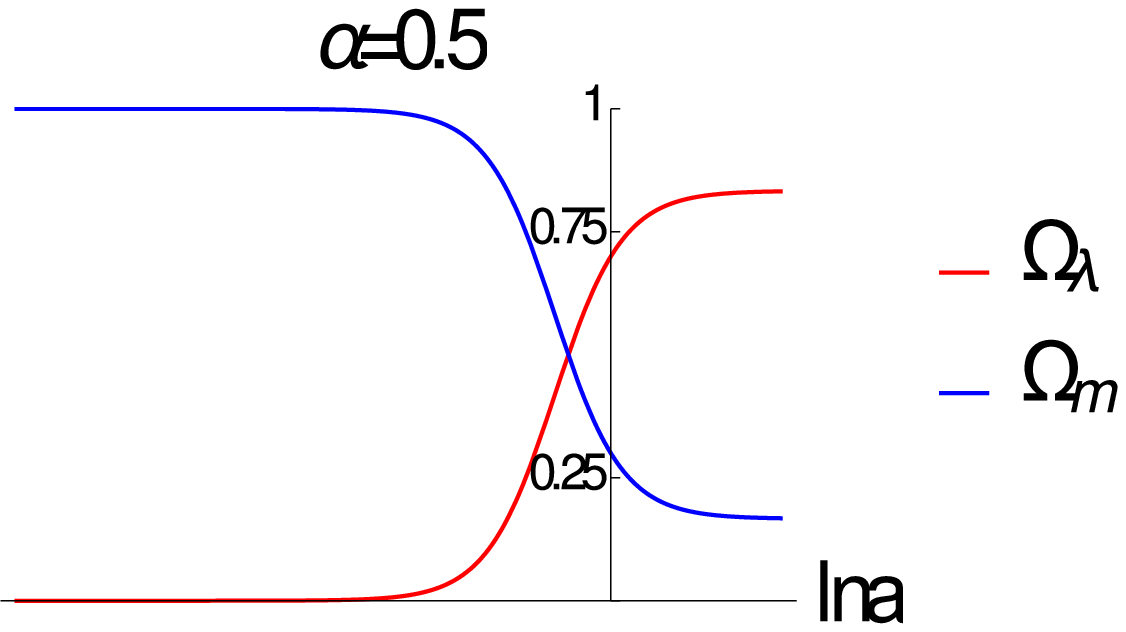}
\hspace{0.1cm}
\small{(d)}\includegraphics[width=7.8cm,height=5cm,clip]{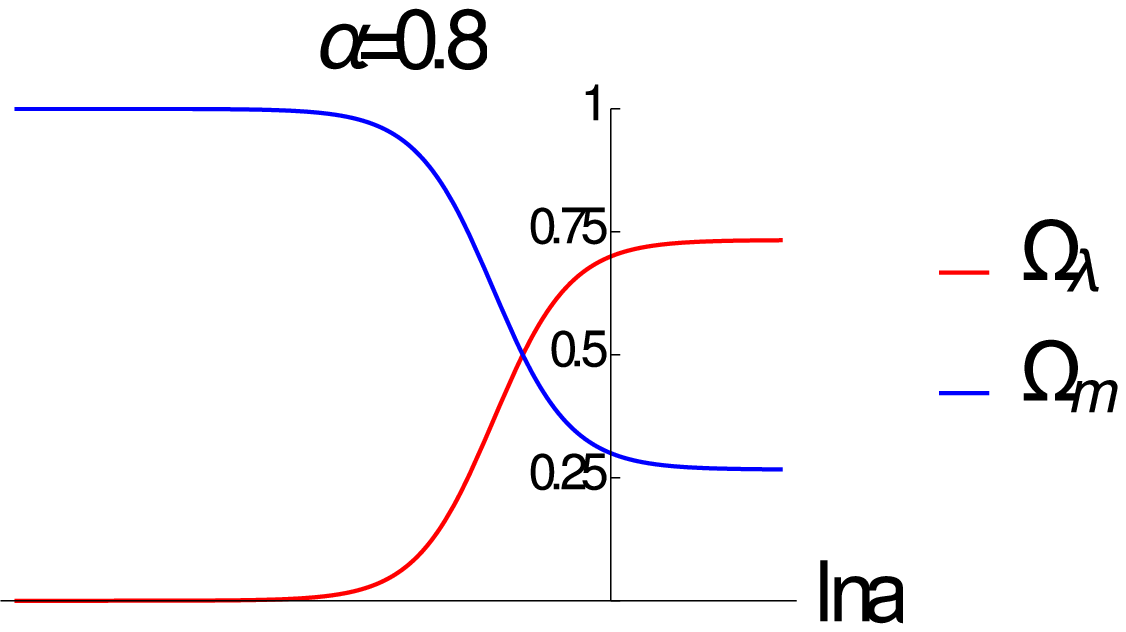}
\end{minipage}
\caption{Plots of $\Omega_{\lambda}$ and $\Omega_m$ as a function of $\eta~(=$ ln$~a)$ for different values coupling parameter $\alpha$. The vertical axis represents the present cosmological time.}
\label{fig:variation}
\end{figure}

\begin{equation}
	\dfrac{\textrm{d}\mathcal{X}}{\textrm{d}\eta}=\dfrac{1}{H}\left[\dfrac{8\pi G}{3H^2}\dot{\rho }_\lambda-2\dfrac{8\pi G}{3H^3}\rho_{\lambda }\dot{H}\right]\label{eq:lna-x-variation1},
\end{equation}
\begin{equation}
	\dfrac{\textrm{d}\mathcal{Y}}{\textrm{d}\eta}=\dfrac{1}{H}\left[\dfrac{8\pi G}{3H^2}\dot{\rho }_m-2\dfrac{8\pi G}{3H^3}\rho_{m}\dot{H}\right]\label{eq:lna-x-variation2}.
\end{equation}
From Eq.(\ref{eq:conser3}), we get
\begin{equation}
\dot{\rho}_\lambda=-\alpha H\rho_\lambda,\;\;\;\;\;\; \dot{\rho}_m=\alpha H\rho_{\lambda}-3 H\rho_{m}\label{eq:energy-density-m,lambda},
\end{equation}
and from Eq.(\ref{eq:friedmann-eqn}) we get
\begin{equation}
	\dfrac{\dot{H}}{H^2}=-\dfrac{3}{2}\dfrac{8\pi G}{3H^2}\rho_{m},\;\;\;\;\;\; \mathcal{X}+\mathcal{Y}=1\label{eq:modified-friedmann}.
\end{equation}
So using, Eq.(\ref{eq:energy-density-m,lambda}), Eq (\ref{eq:modified-friedmann}) in Eq.(\ref{eq:lna-x-variation1}), Eq(\ref{eq:lna-x-variation2}), we get the following set of equations
\begin{equation}
	\dfrac{\textrm{d}\mathcal{X}}{\textrm{d}\eta}+3\mathcal{X}^2+(\alpha-3)\mathcal{X}=0\label{eq:plot-eqn1},
\end{equation}
\begin{equation}
	\dfrac{\textrm{d}\mathcal{Y}}{\textrm{d}\eta}-3\mathcal{Y}^2+(\alpha+3)\mathcal{Y}-\alpha=0\label{eq:plot-eqn2}.
\end{equation}
 
 
\begin{figure}
 \includegraphics[scale=0.99]{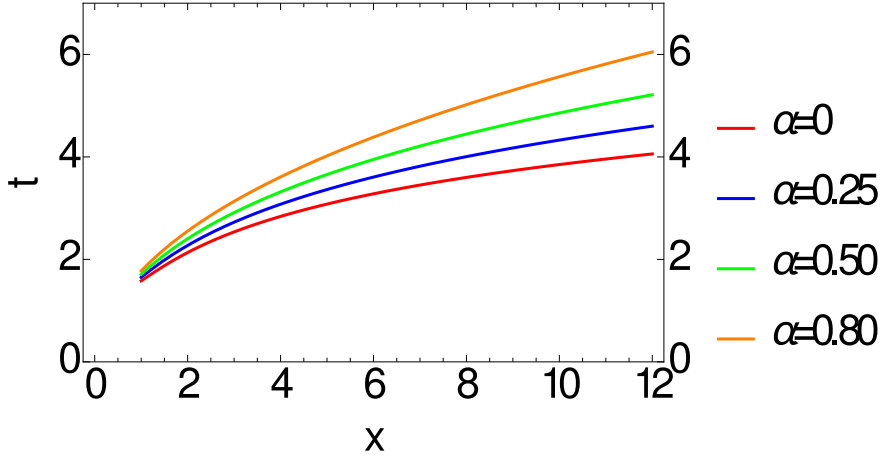}
 \caption{Graphical variation of Eq.(\ref{eq:hypergeometric}) for $\Omega_{\lambda}^0=0.7$ and $\Omega_m^0=0.3$ where cosmic time (in the unit of $H_0^{-1}$) is plotted against normalized dimensionless scale factor $x$ in presence and absence of coupling.}
	\label{fig:tvsx}
\end{figure}



\begin{figure}
 \includegraphics[scale=1]{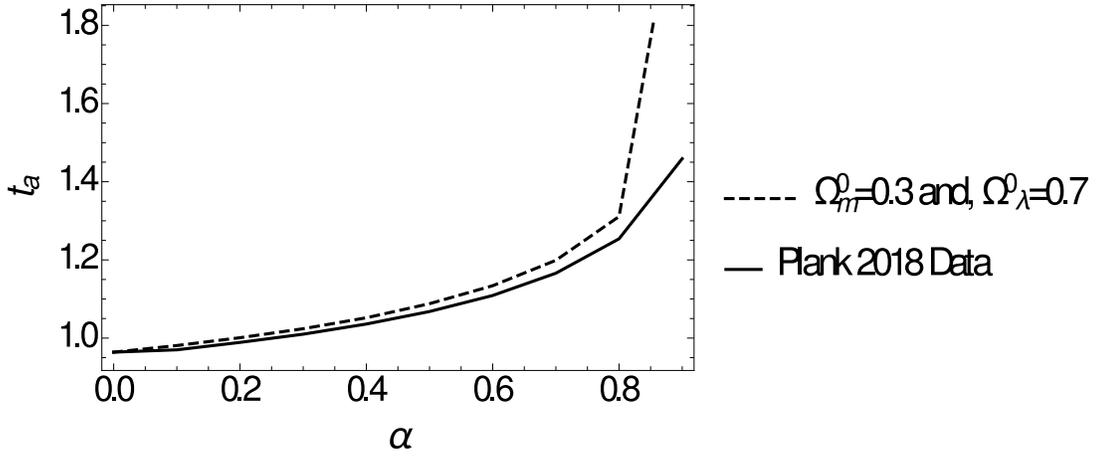}
 \caption{Graphical variation of Table [\ref{tab:tab1}]. Here we observe the evolution of Age of the Universe with the coupling constant $\alpha$.}
	\label{fig:var2}
\end{figure}

We have plotted Eq.(\ref{eq:plot-eqn1}), (\ref{eq:plot-eqn2}) in Fig.[\ref{fig:variation}]. In Fig.[\ref{fig:tvsx}] we have observed the variation of normalized cosmic time $t$ with scale factor $x$ (Eq.(\ref{eq:hypergeometric})), and
in Fig[\ref{fig:var2}] we plotted the data of Table [\ref{tab:tab1}] which shows the age of universe for different value of coupling constant $\alpha$.



\section{Conclusion}
In this article we studied the evolution of scale factor by considering the two interacting components (matter and dark energy) in the spatially flat universe.  We considered the spatially homogeneous tachyonic scalar field as a candidate of dark energy. In the interacting dark energy model, the two components are mutually coupled via coupling parameter $Q$, and a transfer of energy between the two components is possible. During the interaction, the individual components can violate the energy conservation, but overall energy is conserved. In the absence of a fundamental theory of dark sector the choice of coupling parameter is purely phenomenological. We obtained the age of the universe in interacting dark energy model and presented in Table [\ref{tab:tab1}] for different value of coupling constant. We found that with the increase in the value of coupling constant ($\alpha$), the age of the universe is also increasing. But as we further increased the value of coupling constant (beyond 1), the age of the universe turns out to be imaginary which is a non-physical situation. This puts an upper bound to the coupling constant $\alpha$, and it should be less than 1. We plotted the normalized energy density of dark energy ($\Omega_\lambda$) and matter ($\Omega_m$) as a function of $\text{ln}~ a$ in Fig.[\ref{fig:variation}]. Fig.[\ref{fig:tvsx}] showed the relationship between the cosmic time and the normalized dimensionless scale factor $x$. And in Fig.[\ref{fig:var2}], we plotted the data of Table[\ref{tab:tab1}] which shows the variation of the age of the universe with the coupling constant $\alpha$.


\section*{Acknowledgment}
The authors are thankful to reviewers for their useful comments and suggestions.


\bibliography{CosmologyLiterature.bib}

\begin{thebibliography}{10}

\bibitem{Riess_1998}
A.~G. Riess, A.~V. Filippenko, P.~Challis, A.~Clocchiatti, A.~Diercks, P.~M.
  Garnavich, R.~L. Gilliland, C.~J. Hogan, S.~Jha, R.~P. Kirshner, and et~al.,
  ``Observational evidence from supernovae for an accelerating universe and a
  cosmological constant,'' {\em The Astronomical Journal}, vol.~116,
  p.~1009–1038, Sep 1998.

\bibitem{Perlmutter_1999}
S.~Perlmutter, G.~Aldering, G.~Goldhaber, R.~A. Knop, P.~Nugent, P.~G. Castro,
  S.~Deustua, S.~Fabbro, A.~Goobar, D.~E. Groom, and et~al., ``Measurements of
  Ω and Λ from 42 high‐redshift supernovae,'' {\em The Astrophysical
  Journal}, vol.~517, p.~565–586, Jun 1999.

\bibitem{Chimento_2010}
L.~P. Chimento, ``Linear and nonlinear interactions in the dark sector,'' {\em
  Physical Review D}, vol.~81, p.~043525, Feb 2010.

\bibitem{Chimento_2008}
L.~P. Chimento, M.~Forte, and G.~M. Kremer, ``Cosmological model with
  interactions in the dark sector,'' {\em General Relativity and Gravitation},
  vol.~41, p.~1125–1137, Sep 2008.

\bibitem{Bertolami_2012}
O.~Bertolami, P.~Carrilho, and J.~Páramos, ``Two-scalar-field model for the
  interaction of dark energy and dark matter,'' {\em Physical Review D},
  vol.~86, p.~103522, Nov 2012.

\bibitem{Wang_2007}
B.~Wang, J.~Zang, C.-Y. Lin, E.~Abdalla, and S.~Micheletti, ``Interacting dark
  energy and dark matter: Observational constraints from cosmological
  parameters,'' {\em Nuclear Physics B}, vol.~778, p.~69–84, Aug 2007.

\bibitem{lu2012investigate}
J.~Lu, Y.~Wu, Y.~Jin, and Y.~Wang, ``Investigate the interaction between dark
  matter and dark energy,'' {\em Results Phys.}, vol.~2, pp.~14--21, 2012,
  1203.4905.

\bibitem{Farajollahi_2012}
H.~Farajollahi, A.~Ravanpak, and G.~Fadakar, ``Interacting agegraphic dark
  energy model in tachyon cosmology coupled to matter,'' {\em Physics Letters
  B}, vol.~711, p.~225–231, May 2012.

\bibitem{Zimdahl_2012}
W.~Zimdahl, ``Models of interacting dark energy,'' {\em AIP Conf. Proc.},
  vol.~1471, pp.~51--56, 2012, 1204.5892.

\bibitem{Yang_2018}
W.~Yang, S.~Pan, E.~D. Valentino, R.~C. Nunes, S.~Vagnozzi, and D.~F. Mota,
  ``Tale of stable interacting dark energy, observational signatures, and the
  h0 tension,'' {\em Journal of Cosmology and Astroparticle Physics},
  vol.~2018, p.~019–019, Sep 2018.

\bibitem{Cao_2011}
S.~Cao, N.~Liang, and Z.-H. Zhu, ``Testing the phenomenological interacting
  dark energy with observational h(z) data,'' {\em Monthly Notices of the Royal
  Astronomical Society}, vol.~416, p.~1099–1104, Jul 2011.

\bibitem{Di_Valentino_2020}
E.~Di~Valentino, A.~Melchiorri, O.~Mena, and S.~Vagnozzi, ``Nonminimal dark
  sector physics and cosmological tensions,'' {\em Physical Review D},
  vol.~101, p.~063502, Mar 2020.

\bibitem{Verma_2012}
M.~M. Verma and S.~D. Pathak, ``A tachyonic scalar field with mutually
  interacting components,'' {\em International Journal of Theoretical Physics},
  vol.~51, p.~2370–2379, Mar 2012.

\bibitem{Verma_2013}
M.~M. Verma and S.~D. Pathak, ``Shifted cosmological parameter and shifted dust
  matter in a two-phase tachyonic field universe,'' {\em Astrophysics and Space
  Science}, vol.~344, p.~505–512, Jan 2013.

\bibitem{V_liviita_2010}
J.~Väliviita, R.~Maartens, and E.~Majerotto, ``Observational constraints on an
  interacting dark energy model,'' {\em Monthly Notices of the Royal
  Astronomical Society}, vol.~402, p.~2355–2368, Mar 2010.

\bibitem{Pan_2018}
S.~Pan, A.~Mukherjee, and N.~Banerjee, ``Astronomical bounds on a cosmological
  model allowing a general interaction in the dark sector,'' {\em Monthly
  Notices of the Royal Astronomical Society}, vol.~477, p.~1189–1205, Mar
  2018.

\bibitem{Amendola_2018}
L.~Amendola, J.~Rubio, and C.~Wetterich, ``Primordial black holes from fifth
  forces,'' {\em Physical Review D}, vol.~97, p.~081302, Apr 2018.

\bibitem{B_gu__2019}
D.~Bégué, C.~Stahl, and S.-S. Xue, ``A model of interacting dark fluids
  tested with supernovae and baryon acoustic oscillations data,'' {\em Nuclear
  Physics B}, vol.~940, p.~312–320, Mar 2019.

\bibitem{Pan_2019}
S.~Pan, W.~Yang, C.~Singha, and E.~N. Saridakis, ``Observational constraints on
  sign-changeable interaction models and alleviation of the h0 tension,'' {\em
  Physical Review D}, vol.~100, p.~083539, Oct 2019.

\bibitem{Papagiannopoulos_2020}
G.~Papagiannopoulos, P.~Tsiapi, S.~Basilakos, and A.~Paliathanasis, ``Dynamics
  and cosmological evolution in $\lambda $ Λ-varying cosmology,'' {\em The
  European Physical Journal C}, vol.~80, p.~55, Jan 2020.

\bibitem{Savastano_2019}
S.~Savastano, L.~Amendola, J.~Rubio, and C.~Wetterich, ``Primordial dark matter
  halos from fifth forces,'' {\em Physical Review D}, vol.~100, p.~083518, Oct
  2019.

\bibitem{von_Marttens_2019}
R.~von Marttens, L.~Casarini, D.~Mota, and W.~Zimdahl, ``Cosmological
  constraints on parametrized interacting dark energy,'' {\em Physics of the
  Dark Universe}, vol.~23, p.~100248, Jan 2019.

\bibitem{yang2019reconstructing}
W.~Yang, N.~Banerjee, A.~Paliathanasis, and S.~Pan, ``Reconstructing the dark
  matter and dark energy interaction scenarios from observations,'' {\em Phys.
  Dark Univ.}, vol.~26, p.~100383, 2019, 1812.06854.

\bibitem{Asghari_2019}
M.~Asghari, J.~B. Jiménez, S.~Khosravi, and D.~F. Mota, ``On structure
  formation from a small-scales-interacting dark sector,'' {\em Journal of
  Cosmology and Astroparticle Physics}, vol.~2019, p.~042–042, Apr 2019.

\bibitem{Sen_2002}
A.~Sen, ``Rolling tachyon,'' {\em Journal of High Energy Physics}, vol.~2002,
  p.~048–048, Apr 2002.

\bibitem{Sen1_2002}
A.~Sen, ``Tachyon matter,'' {\em Journal of High Energy Physics}, vol.~2002,
  p.~065–065, Jul 2002.

\bibitem{Sen2_2002}
A.~Sen, ``Field theory of tachyon matter,'' {\em Modern Physics Letters A},
  vol.~17, p.~1797–1804, Sep 2002.

\bibitem{pathak2016thermodynamics}
S.~D. Pathak, M.~M. Verma, and S.~Li, ``Thermodynamics of interacting tachyonic
  scalar field,'' in {\em National Conference on Current Issues in Cosmology,
  Astrophysics and High Energy Physics}, (Dibrugarh, India), pp.~73--77,
  Dibrugarh Univ., 2016, 1612.00860.

\bibitem{Amendola_2003}
L.~Amendola, C.~Quercellini, D.~Tocchini-Valentini, and A.~Pasqui, ``Cosmic
  microwave background as a gravity probe,'' {\em The Astrophysical Journal},
  vol.~583, p.~L53–L56, Feb 2003.

\bibitem{Berger_2008}
M.~S. Berger and H.~Shojaei, ``Possible equilibria of interacting dark energy
  models,'' {\em Physical Review D}, vol.~77, p.~123504, Jun 2008.

\bibitem{Verma_2014}
M.~M. Verma and S.~D. Pathak, ``The bicep2 data and a single higgs-like
  interacting scalar field,'' {\em International Journal of Modern Physics D},
  vol.~23, p.~1450075, Aug 2014.

\bibitem{Shahalam_2015}
M.~Shahalam, S.~D. Pathak, M.~M. Verma, M.~Y. Khlopov, and R.~Myrzakulov,
  ``Dynamics of interacting quintessence,'' {\em The European Physical Journal
  C}, vol.~75, p.~395, Aug 2015.

\bibitem{Pav_n_2008}
D.~Pavón and B.~Wang, ``Le châtelier–braun principle in cosmological
  physics,'' {\em General Relativity and Gravitation}, vol.~41, p.~1–5, Jun
  2008.

\bibitem{Akash_2020}
K.~Akash, S.~D. Pathak, and R.~K. Dubey, ``Dynamical role of scalar fields in k
  = 0 universe,'' {\em Journal of Physics: Conference Series}, vol.~1531,
  p.~012086, may 2020.

\bibitem{Hypergeometric_book}
K.~S. Rao and V.~Lakshminarayanan, {\em Generalized Hypergeometric Functions}.
\newblock 2053-2563, IOP Publishing, 2018.

\bibitem{planck2020}
N.~Aghanim, Y.~Akrami, F.~Arroja, M.~Ashdown, J.~Aumont, C.~Baccigalupi,
  M.~Ballardini, A.~J. Banday, R.~B. Barreiro, and et~al., ``Planck 2018
  results,'' {\em Astronomy and Astrophysics}, vol.~641, p.~A1, Sep 2020.

\bibitem{Condon_2018}
J.~J. Condon and A.~M. Matthews, ``Λcdm cosmology for astronomers,'' {\em
  Publications of the Astronomical Society of the Pacific}, vol.~130,
  p.~073001, jun 2018.

\bibitem{Bahamonde_2018}
S.~Bahamonde, C.~G. Böhmer, S.~Carloni, E.~J. Copeland, W.~Fang, and
  N.~Tamanini, ``Dynamical systems applied to cosmology: Dark energy and
  modified gravity,'' {\em Physics Reports}, vol.~775-777, p.~1–122, Nov
  2018.

\end{thebibliography}
\bibliographystyle{hieeetr}
\end{document}